# An infinite-server queueing model $MMAP_k|G_k|\infty$ in semi-Markov random environment with marked MAP arrival and subject to catastrophes

K. Kerobyan [1], R. Covington[1], R. Kerobyan [2], K. Enakoutsa[1]

[1] khanik.kerobyan@csun.edu , [2] rkerobya@ucsd.edu
[1]*California State University Northridge, CSUN, Northridge, USA;*
[2]*University of California San Diego, UCSD, San Diego, USA*

In the present paper the infinite-server $MMAP_k|G_k|\infty$ queueing model with random resource vector of customers, marked MAP arrival and semi-Markov (SM) arrival of catastrophes is considered. The joint generating functions (PGF) of transient and stationary distributions of number of busy servers and numbers of different types served customers, as well as Laplace transformations (LT) of joint distributions of total accumulated resources in the model at moment *t* and total accumulated resources of served customers during time interval $[0,t)$ are found. The basic differential and renewal equations for transient and stationary PGF of queue sizes of customers are found.

**Key words:** marked MAP, infinite-server queue model, catastrophe, resource vector

**Introduction**

From the time of Erlang's pioneering research [1], queueing models with many servers, in particular, the infinite-server models, have been widely used for modelling and performance evaluation of wired and wireless computer and telecommunication networks [2]. As shown by large number of measurements the traffic of modern computer networks has self-similar nature and can be characterized by the heterogeneousness, the non-stationarity, the burstiness, and the correlations [3]. Network traffics in queue are generally described by traffic models based on finite Markovian Processes: Markov Arrival Process (MAP), Batch MAP (BMAP), Marked MAP (MMAP) and their generalizations [3,4]. The MMAP and MAP arrivals properties and their applications are presented in [4,5,6] and are not duplicated here.

The $BMAP|G|\infty$ model and some of its predictions studied by [7,8]. By using semi-Markov processes (SMP), matrix analytic methods the PGF of the number of busy servers and its moments are considered. The models with phase type arrival $PH|G|\infty$ and Markov modulated arrival $N|G|\infty$ are considered in [9,10]. The queueing model $M_k|M_k|\infty$ with correlated k heterogeneous customers in a batch and exponential service time is studied by [11]. The joint PGF of the number of type *k* customers being served in the system is derived explicitly by solving partial differential equations. The generalization of this model for general service time $M_k|G_k|\infty$ Poisson arrival of customers and $BM_k|G_k|\infty$ with *k* correlated heterogeneous customers in a batch is considered in [12,13]. In steady state, the joint PGF of queue length of customers by using collective mark method CMM and conditional expectations is derived in [13]. The model $MAP_k|G_k|\infty$ with the structured batch arrival of *k* types of customers is considered in [14]. In steady state, the differential equations for PGF of the number of busy server and its solution are obtained. In steady state, the differential equations for PGF of queue length and its moments are obtained. The first and second order asymptotes of queue length for the models $MAP|G|\infty$, $MMPP|G|\infty$, $G|G|\infty$ based on supplementary variable method are studied in [15].

To evaluate the impact of network environment on networks performance metrics the infinite-server models in the random enviroment (RE) are applied. The queue size distribution of the model $M|G|\infty$ in semi-Markov (SM) environment is studied in [16-19]. By the method of supplementary variable and the original method of dynamic screening first and second order asymptotes of queue size distribution are obtained [16]. The stochastic decomposition formula for queue length distribution is obtained in [17]. The queue M|G|∞ in random environment with clearing mechanism is studied in [17,18]. The environmental clearing process is modeled by an *m*-state irreducible SMP. The transient and steady-state queue length distributions by using renewal arguments are obtained. The MMAP$_k$|G$_k$|∞ queue in SM environment and catastrophes is studied in [20]. The PGFs of joint distributions of queue size and number of served customers by using renewal arguments and differential equations are found. The of infinite-server queue MMAP$_k$(t)|G$_k$|∞ with Poisson stream catastrophes and nonhomogeneous marked MAP arrival of customers is studied in [21]. The PGF of queue length of different types of customers is obtained. The model M|M|∞ with disasters in steady-state is studied in [22,23].

In many applications of queueing models such as computer and communication networks and systems, production systems, transportation systems, economics, finances and insurance systems the customers characterize by vector of requesting resources which components can be determinic or random quantities. For instance one component can describe number of servers necessary to serve the customer, second – amount of time necessary to serve the customer, next – the volume of space to save the customer and so on. Also the customers must have some features necessary to be accepted by system. Some components of resource vectors can be discrete (number of servers, amount of parts) and others can be continuous (space-volume to save, amount of finances, power).

Despite importance of resource models of customers in queuing theory, there are very few works devoted to research such kind of models, see for example [20,24-29].

The main methods to study the infinite-server queues are: supplementary variables method [30], the method which based on conditional expectations [30,31], and collective marks method (CMM) method [30,32,33]. The last method is also called "supplementary event" [30] or "catastrophes" method [32, 33] and has been used successfully for queue models with priorities [33]. CMM have been used for infinite-server models in [13] for BM$_k$|G$_k$|∞ queue with Poisson arrival of batches and in [34] for M|SM|∞ queue. In [34] the method is mentioned but does not used to obtain some results.

In this paper we consider some generalizations of [17,20,24] results for infinite-server MMAP$_k$|G$_k$|∞ queue in random environment and catastrophes. The joint PGF of transient and stationary distributions of number of busy servers and numbers of served customers, as well as the LT of joint distributions of total accumulated resource in the model and total accumulated resource vectors of served customers during time interval $[0,t)$ are found. The renewal equations for transient and stationary PGF of queue sizes of different types of customers are found as well. All results are obtained using CMM and renewal process methods.

### 1. Model Description

We consider an infinite-server MMAP|G|∞ queue model in random environment (RE) with *K* types of customers and catastrophes. The RE operates according to stationary, irreducible semi-Markov process (SMP) $\xi(t)$, $t \geq 0$ with finite state space $S = \{1, 2, ..., k\}$. The SMP is given by the vector of initial distribution $p^0 = \{p_i^0, i \in S\}$ and SM matrix $Q(t) = \|Q_{ij}(t)\|$, $t \geq 0, i, j \in S$. Customers arrive according to homogeneous marked MAP which is given by the sequence of characteristic matrices $\{D_0, D_h, \boldsymbol{h} \in C^0\}$. Here $C^0$ is a finite or counting set of arriving batches, $\boldsymbol{h} = (h_1, h_2, ..., h_K)$, $\boldsymbol{h} \in C^0$, and $h_r$ is a number of type *r* customers, $0 \leq h_r$, $1 \leq r \leq K$ in a batch. $D_0$ is a non-singular matrix with negative diagonal elements and $D_h$ are non-negative matrices, $\boldsymbol{h} \in C^0$. The phase process (PP) $J(t)$ of MMAP is an irreducible Markov process (MP) with

generating matrix (GM) $D$ and finite set of states $E$. $D$ is a matrix of $m \times m$ size.

$$D = D_0 + \sum_{h \in C^0} D_h, \quad De = 0, \quad \pi D = 0, \quad \pi e = 1,$$

where $e$ is the column vector with all components one, $\pi$ is the vector of stationary distribution $\pi = (\pi_1, ..., \pi_m)$ of PP $J(t)$.

The service of arriving customers begins immediately. Let the random variable (r.v.) $\gamma_r$ be a service time of type $r$ customers, and $\gamma = (\gamma_1, \gamma_2, ..., \gamma_K)$ is a vector of service times. Components of $\gamma$ are independent, identically distributed (i.i.d.) r.v.s which depend on type of the customer and state of environmental SMP. R.v. $\gamma_r$ has $B_r(t) = P(\gamma_r < t)$ general distribution and finite mean value $\bar{\gamma}_r, 1 \leq r \leq K$.

Each arriving customer is characterized by $k$-dimensional random volume (resource) vector $\zeta_r = (\zeta_{1r}, ..., \zeta_{kr})$ and each departing customer is characterized by random vector $\sigma_r = (\sigma_{1r}, ..., \sigma_{kr})$ with non-negative components $1 \leq r \leq K$.

Let $C_r(x) = P(\zeta_{1r} \leq x_1, ..., \zeta_{kr} \leq x_k)$ and $G_r(x) = P(\sigma_{1r} \leq x_1, ..., \sigma_{kr} \leq x_k)$ are the joint distributions of resource vectors $\zeta_r$ and $\sigma_r$, where $x = (x_1, ..., x_k)$. We assume that the service time vector $\gamma$ and the resource vectors $\zeta = (\zeta_1, ..., \zeta_K)$, $\sigma = (\sigma_1, ..., \sigma_K)$ are mutually independent.

When SMP $\xi(t), t \geq 0$ jumps from state $i$ to the state $r$ all customers in the model are instantly flashed out and the model jumps into empty state. Let consider the related with MMAP counting processes $N(t), N_s(t), M(t): N(t) = (N_1(t), .., N_K(t))$, $N_s(t) = (N_{1s}(t), .., N_{Ks}(t))$, $M(t) = (M_1(t), .., M_K(t))$, where $N_r(t)$ and $M_r(t)$ are the number of type $r$ customers arriving and serving in time interval $[0,t)$, and $N_{rs}(t)$ is number of customers being in service at moment $t$. Let $\boldsymbol{\beta}(t) = (\beta_1(t), ..., \beta_K(t))$ and $\boldsymbol{\alpha}(t) = (\alpha_1(t), .., \alpha_K(t))$ be the vectors of total resource served during interval $[0,t)$, and accumulated in the model at moment $t$. The components of $\boldsymbol{\beta}(t)$, $\boldsymbol{\alpha}(t)$, $N_s(t)$ and $M(t)$ vectors are defined as:

$$\beta_r(t) = \sum_{i=1}^{M_r(t)} \sigma_{ri}, \quad \alpha_r(t) = \sum_{i=1}^{N_r(t)} \zeta_{ri}, \quad N_{rs}(t) = \sum_{h \in C^0} N_{h_r}^s(t) \quad M_r(t) = \sum_{h \in C^0} M_{h_r}(t) \quad r = 1, 2, .., K.$$

Suppose that at initial time $t = 0$ model is empty, $N(0) = 0$, $M(0) = 0$, $\alpha(0) = 0$, $\beta(0) = 0$.

## 2. The counting process.

Let consider the counting process (CP) $\{N(t), J(t); t \geq 0\}$ with matrix $P(n,t)$ of transition probabilities: $P(n,t) = \| p_{ij}(n,t) \|$, $p_{ij}(n,t) = P(N(t) = n, J(t) = j | J(0) = i)$, $1 \leq i, j \leq m$, where $n = (n_1, ..., n_k)$: $n_i$ are non-negative integers. Let define the following generating functions (GF) $D(z)$, $P(z,t)$

$$D(z) = D_0 + \sum_{h \in C^0} z^h D_h, \quad |z_r| \leq 1, \quad 0 \leq r \leq K, \quad P(z,t) = \sum_{n \geq 0} z^n P(n,t),$$

where $z = (z_1, z_2, ..., z_k)$ and $z^h = (z_1^{h_1}, z_2^{h_2}, ..., z_k^{h_k})$.

**Theorem 1.** The PGF of counting process $\{N(t), J(t); t \geq 0\}$ $P(z,t)$ satisfies the *basic differential equation*

$$\frac{\partial}{\partial t} P(z,t) = D(z) P(z,t), \quad |z| \leq 1, \tag{1}$$

with initial conditions $P(z,0) = 1$.
The solution of differential equation (1) is given by

$$P(z,t) = \exp\{D(z)t\}. \tag{2}$$

**Proof.** The transition probabilities $\{P(n,t), n \geq 0\}$ of CP $N(t)$ satisfy the following Kolmogorov backward differential equations

$$\frac{d}{dt} P(n,t) = P(n,t) D_0 + \sum_{h \leq n, h \in C^0} P(n-h, t) D_h, \quad n \geq 0,$$

with initial condition $P_i(n,0) = 0, n > 0$, $P_i(0,0) = 1, i = 1, 2, .., k$.

Pre-multiplying each equation by corresponding $z^n$ after summation we get differential equations for PDF $P(z,t)$. The solution of this equation in matrix exponential form is given by (2).

**Remark.** The proof of (1) by using the CMM is routine and tedious. To show the main procedures of proof let consider related with MAP CP $\{N(t), J(t); t \geq 0\}$. Suppose that MAP is given by two characteristic matrices $D_0, D_1$. From (1) it follows that PGF $P(z,t)$ of counting process $\{N(t), J(t); t \geq 0\}$ is definite as $P(z,t) = \exp\{D(z)t\}$, where $D(z) = D_0 + D_1 z$. The PGF $P(z,t)$ can be written as

$$P(z,t) = \sum_{n=0}^{\infty} e^{D_0 t} \frac{(tD_1)^n}{n!} z^n.$$

Let consider the following coloring scheme of customers. Each arriving customer we mark *red* with probability $z$ and *blue* with $1-z$ probability. Then $P(z,t)$ is the probability that no *blue* customers arrive in $[0,t)$. Let consider the events "$n$ customers arrive in interval $[0,t)$ and all $n$ customers are *red*". The probability of this event is $e^{D_0 t} \frac{(tD_1)^n}{n!} z^n$. By total probability rule we find the probability "all customers arriving in interval $[0,t)$ are no *blue*".

By Theorem 1, we can find moments of CP $\{N(t), J(t); t \geq 0\}$. For example, the mean $E[N_h(t)]$ and variance $Var[N_h(t)]$ of CP $N(t)$ can be found explicitly

$$E[N_h(t)] = \lambda_h t + \theta(e^{Dt} - I)(D - e\pi)^{-1} D_h e, \quad t \geq 0,$$

where $\theta$ is a initial distribution of PP $J(t)$, $\lambda_h = \pi D_h e$ is the stationary arrival rate of type $h$ batches.

$$Var[N_h(t)] = [\lambda_h - 2\lambda_h^2 - 2\pi D_h (D - e\pi)^{-1} D_h e] t + 2\pi D_h (D - e\pi)^{-1} (e^{Dt} - I)(D - e\pi)^{-1} D_h e, \quad t \geq 0.$$

If as an initial distribution use vector $\pi$ then for stationary arrival rate of customers we get $E[N_h(t)] = \lambda_h t$. The stationary arrival rate of all batches of customers is given by $\lambda = \sum_h \pi D_h e$.

The stationary arrival rates of type $r$ customers and all customers regardless of type arriving in $[0,t)$ are given by

$$E[N_r(t)] = \lambda_r t, \quad E[N(t)] = \lambda t = \sum_{r=1}^{K} \lambda_r t,$$

where $\lambda_r = \sum_{n=1}^{\infty} n \sum_{h \in C^0, h_r = n} \pi D_h e$.

### 3. Thinning MMAP

Let consider the following Bernoulli thinning process of MMAP. Each type $r$ customer which arrives at moment $t$ can join to main stream by probability $p_r(t)$ and can be ignored by probability $1 - p_r(t)$. It can be shown (see e.g. [5,34]), that the main stream is an MMAP process with characteristic matrices $\{D_0(t), D_h(t)\}$.

**Lemma.** The thinned process is a MMAP which counting process has matrix GF $D_T(z,t)$ and PGF $P_T(z,t)$, which are defined as follow

$$D_T(z,t) = \sum_{h_1 \geq 0} \sum_{\substack{h_2 \geq 0 \\ h_1 + h_2 + \ldots h_k \geq 1}} \ldots \sum_{h_r \geq 0} D_h \prod_{r=1}^{k} [1 - p_r(t) + z_r p_r(t)]^{h_r}, \quad (3)$$

$$P_T(z,t) = \exp\{\int_0^t D_T(z,x) dx\}. \quad (4)$$

The proof of lemma based on Theorem 1, MMAP properties and Bernoulli thinning operation properties [5,34]. Let just interpret the GM $D_T(z,t)$ by CMM. Let each type $r$ customer of thinned stream marks *red* with probability $z_r$ and marks *blue* with probability $1 - z_r$. Then left side of (3) is the total rate of not *blue* customers at moment $t$. Right side means, the batch $\mathbf{h} = (h_1, h_2, ..., h_k)$ arrives with rate $D_h$ at moment $t$. In this batch each type $r$ customer is *red* with probability $1 - p_r(t) + z_r p_r(t)$ and all $h_r$ type $r$ customers in this batch are *red* with probability $[1 - p_r(t) + z_r p_r(t)]^{h_r}$. The probability that in arriving batch $\mathbf{h}$ are not any *blue* customers is $\prod_{r=1}^{k}[1 - p_r(t) + z_r p_r(t)]^{h_r}$. Then the product $D_h \prod_{r=1}^{k}[1 - p_r(t) + z_r p_r(t)]^{h_r}$ is a rate of *red* batch of size $\mathbf{h}$ arriving at moment $t$. Taking the sum over all possible batches arriving at moment $t$ finally we obtain $D_T(z,t)$ the rate of *not blue* customers arriving at moment $t$.

The subject of our interest is the joint distribution
$$P(n, m, x, y, t) = P(N_s(t) = n, M(t) = m, \alpha(t) \leq x, \beta(t) \leq y).$$

## 4. Model Analysis

Let suppose that the environmental SMP $\xi(t)$, $t \geq 0$ is in state $i \in S$ and consider the dynamic of the the model during time interval $[u,t)$. Each type $r$ customer arriving at moment $u$ will be in service at moment $t$ by probability $1 - B_{ri}(t-u)$ and will finish its service before moment $t$ by probability $B_{ri}(t-u)$.

Let $A_{jk}^i(n, m, x, y, u, t)$ be the joint probability that $n$ customers are in service at moment $t$, and $m$ customers are already served in $[0,t)$, total accumulated resources in the model at moment $t$ is $\alpha(t) \leq x$ and total accumulated served resource during interval $[0,t)$ is $\beta(t) \leq y$, PP $J(u)$ is in phase $j \in E$ under condition that at initial moment $t = 0$ the model was empty, and PP $J(0)$ was in phase $k \in E$:

$A_{jk}^i(n, m, x, y, u, t) = P(N_i^s(u,t) = n, M_i(u,t) = m, \alpha(t) \leq x, \beta(t) \leq y, J(u) = j \mid N_i^s(0) = 0, M_i(0) = 0, J(0) = k)$.

Let $\tilde{A}^i(z_1, z_2, s_1, s_2, u, t) = \| \tilde{A}_{jk}^i(z_1, z_2, s_1, s_2, u, t) \|$ be the matrix which elements are Laplace - Stieltjes transformation (LST) and $z$ transformation of $A_{jk}^i(n, m, x, y, u, t)$; $\tilde{F}_{ri}(s_1)$ and $\tilde{G}_{ri}(s_2)$ be the LST of $F_{ri}(x)$ and $G_{ri}(y)$. For homogeneous model we have $\tilde{A}^i(z_1, z_2, s_1, s_2, t) = \tilde{A}^i(z_1, z_2, s_1, s_2, u, t)$, e.g. see [7].

$$\tilde{A}^i(z_1, z_2, s_1, s_2, t) = \sum_{n=0}^{\infty} \sum_{m=0}^{\infty} z_1^n z_2^m \int_0^{\infty} \int_0^{\infty} e^{-s_1 x - s_2 y} A^i(n, m, dx, dy, t), \; |z_1| \leq 1, \; |z_2| \leq 1,$$

$$\tilde{F}_{ri}(s_1) = \int_0^{\infty} e^{-s_1 x} dF_{ri}(x), \quad \tilde{G}_{ri}(s_2) = \int_0^{\infty} e^{-s_2 y} dG_{ri}(y).$$

For the PGF $\tilde{A}^i(z_1, z_2, s_1, s_2, t)$ of the model MMAP$_r$|G$_r$|$\infty$ can be proved the following result (e.g. see [20,35,36]).

**Theorem 2.** The PGF $\tilde{A}^i(z_1, z_2, s_1, s_2, t)$ satisfy the following basic differential and integral equations

$$\tilde{A}^i(z_1, z_2, s_1, s_2, t) = e^{D_0(i)t} + \int_0^t e^{D_0(i)u} \tilde{S}_i(z_1, z_2, s_1, s_2, u) \tilde{A}^i(z_1, z_2, s_1, s_2, t - u) du. \tag{5}$$

$$\frac{\partial}{\partial t} \tilde{A}^i(z_1, z_2, s_1, s_2, t) = [D_0(i) + \tilde{S}_i(z_1, z_2, s_1, s_2, t)] \tilde{A}^i(z_1, z_2, s_1, s_2, t), \; i \in S, \tag{6}$$

with initial conditions $\tilde{A}^i(z_1, z_2, s_1, s_2, 0) = I$.

Here $\tilde{S}_i(z_1, z_2, s_1, s_2, t) = \sum_{h=1}^{\infty} D_{hi} \prod_{r=1}^{K} [z_{r2} \tilde{G}_{ri}(s_2) B_{ri}(t) + z_{r1} \tilde{F}_{ri}(s_1)(1 - B_{ri}(t))]^{h_r}$.

The proof of (5) can be done by using the method of collective marks [32] or renewal arguments [35-37]. For the proof of (5) we can use (1) and (3). Let consider the proof by the CMM. First of all let note that each type $r$ customer which arrivs at moment $u$ will be served up to moment $t$ with probability $B_r(t-u)$ or will be in the model at moment $t$ with probability

$1 - B_r(t-u)$. We mark each served type $r$ customer *red* or *blue* with probabilities $z_{2r}\tilde{G}_r(s_2)$ and $1 - z_{2r}\tilde{G}_r(s_2)$ resp. Alike, we mark each serving in the model type $r$ customer *"red"* or *"blue"* with probabilities $z_{1r}\tilde{F}_r(s_1)$ and $1 - z_{1r}\tilde{F}_r(s_1)$ resp. Then on the left side of equattion (5) $\tilde{A}^i(z_1, z_2, s_1, s_2, t)$ is the probability that at moment $t$ in the model there are not *blue* customers. This event is possible if either the model is free and during interval of time $[0, t)$ there are not any arrivals of customers (with probability $e^{D_0(i)u}$) or at moment $u$ a batch $h$ arrives (with probability $e^{D_0(i)u} D_h(i)du$), all customers in the batch are *red* (with probability $\prod_{r=1}^{K} [z_{r2}\tilde{G}_{ri}(s_2) B_{ri}(u) + z_{r1}\tilde{F}_{ri}(s_1)(1 - B_{ri}(u))]^{h_r}$), and in interval $t - u$ in the model are not any *blue* customers (with probability $\tilde{A}^i(z_1, z_2, s_1, s_2, t - u)$). After using the total probabilities rule we derive the equation (6).

**Theorem 3.** The solution of (5) and (6) is given by
$$\tilde{A}^i(z_1, z_2, s_1, s_2, t) = \exp\{\int_0^t [D_0(i) + \tilde{S}_i(z_1, z_2, s_1, s_2, u)]du\}, \quad |z_1| \leq 1, |z_2| \leq 1. \quad (7)$$

The proof is based on thinning lemma.

When $z_2 = 1, s_2 = 0$ from (7) we obtain the LST of PGF joint distribution number of busy servers and total accumulated resources in the model at moment $t$: $\tilde{\tilde{P}}(z_1, s_1, t, i) = \tilde{A}^i(z_1, 1, s_1, 0, t)$,
$$\tilde{\tilde{P}}(z_1, s_1, t, i) = \exp\{\int_0^t [D_0(i) + \tilde{S}_i(z_1, s_1, u)]du\}, \quad (8)$$
where $\tilde{S}_i(z_1, s_1, t) = \sum_{h=1}^{\infty} D_{hi} \prod_{r=1}^{K} [B_{ri}(t) + z_{r1}\tilde{F}_{ri}(s_1)(1 - B_{ri}(t))]^{h_r}$.

When $z_1 = 1, s_1 = 0$, from (7) we obtain the LST of PGF joint distribution number of served customers and total served resources during interval $[0, t)$:
$$\tilde{W}^i(z_2, s_2, t) = \exp\{\int_0^t [D_0(i) + \tilde{S}_i(z_2, s_2, u)]du\}, \quad (9)$$
where $\tilde{S}_i(z_2, s_2, t) = \sum_{h=1}^{\infty} D_{hi} \prod_{r=1}^{K} [z_{r2}\tilde{G}_{ri}(s_2) B_{ri}(t) + (1 - B_{ri}(t))]^{h_r}$.

If suppose that at time $t = 0$, there are $\mathbf{h}_0 = (h_{01}, h_{02}, ..., h_{0k})$ initial customers in the model then for $\tilde{A}^i(z_1, z_2, s_1, s_2, t)$ we get
$$\tilde{A}^i(z_1, z_2, s_1, s_2, t) = \prod_{r=1}^{K} [z_{r2}\tilde{G}_{ri}(s_2) B_{ri}(t) + (1 - B_{ri}(t))]^{h_{0r}} e^{\int_0^t [D_0(i) + \tilde{S}_i(z_1, z_2, s_1, s_2, u)]du}.$$

From (7) in particular cases can be found PGFs for PH|G|∞ and BMAP|G|∞ models [7,9].

Let consider an infinite-server queue BG$_K$|G$_K$| ∞ with general distributed interarrival time of batches and $K$ types of customers. Let $A(t)$ be the DF of interarrival epochs between batches, and $B_r(t)$ is a service time DF of type $r$ customers. In each inter-arrival epoch by $a(\mathbf{n}) = a(n_1, ..., n_K)$ probability generates a batch $\mathbf{n}$ with $n_1$ customers of type 1, ..., $n_K$ -customers of type $K$, where $\sum_{n_1}\sum_{n_2}...\sum_{n_k} a(n_1, n_2, ..., n_k) = 1$, $a(0, 0, ..., 0) = 0$.

Then for PGF $\tilde{A}(z_1, z_2, s_1, s_2, t)$ of the model BG$_K$|G$_K$| ∞ we can prove the following result.

**Theorem 4.** The PGF $\tilde{A}(z_1, z_2, s_1, s_2, t)$ satisfies the following basic integral equations
$$\tilde{A}(z_1, z_2, s_1, s_2, t) = 1 - A(t) + \int_0^t \tilde{S}(z_1, z_2, s_1, s_2, u) \tilde{A}(z_1, z_2, s_1, s_2, t - u) dA(u), \quad (10)$$
where $\tilde{S}(z_1, z_2, s_1, s_2, u) = \sum_{\substack{\mathbf{n}=0 \\ n_1+n_2+...+n_k=1}}^{\infty} a(\mathbf{n}) \prod_{r=1}^{K} [z_{r2}\tilde{G}_r(s_2) B_r(t) + z_{r1}\tilde{F}_r(s_1)(1 - B_r(t))]^{n_r}$.

**Proof.** Let mark the customers as in case of theorem 2 proof. Then $\tilde{A}(z_1, z_2, s_1, s_2, t)$ in left side of (10) is the probability of event "no *blue* customers was served in the model during interval of time $[0,t)$ no *blue* customer is serving in the model at moment $t$". This event can be realized in two mutually independent ways: either "the first customer arrives into empty model after time $t$" (with probability $1 - A(t)$) or "first batch of customers arrives at moment $u$, $u < t$" (with probability $dA(u)$). This batch includes $a(\mathbf{n})$ customers of different types and "all customers in the batch are *red*" (with probability $\prod_{r=1}^{K} [z_{r2}\tilde{G}_r(s_2)B_r(t) + z_{r1}\tilde{F}_r(s_1)(1 - B_r(t))]^{n_r}$). Then upplying the total probabilities rule we finally get (10).

**Remark.** Let consider the service policy when we do not distinguish the customers of different batches and all customers in the batch serve together as one customer. In this case the MMAP transforms into equivalent MAP with characteristic matrices $D_0 = C$, $D_1 = \sum_{\mathbf{h} \in C^0} D_\mathbf{h}$. Let the r.v. $\gamma_0$ be a service time of batches of customers with general distribution $B_0(t)$ and mean value $\bar{\gamma}_{01}$. Denote by $N(t)$ number of batches which arrive in interval $[0,t)$ and by $J(t)$ the PP of MAP with generator matrix $D = D_0 + D_1$. $P(n,t)$, $n \geq 0$ are $m \times m$ size matrices with elements

$$P_r(n,t,i) = P\{N^s(t) = n, J(t) = i \mid N^s(0) = 0, J(0) = r\}.$$

Where $P_r(n,t,i)$ is a conditional probability of having $n$ batches in service at time $t$ and PP is in state $i$ given that at initial moment model was empty and PP was in state $r$. If $P(z,t)$ be the PGF of $\{P(n,t), n \geq 0\}$, then it satisfies the following differential equation

$$\frac{d}{dt}P(z,t) = P(z,t)[D_0 + D_1(B_0(x) + z(1 - B_0(x)))], \quad |z| \leq 1, \qquad (11)$$

with initial condition $P(z,0) = I$.

The solution of (11) in matrix exponential form is: $P(z,t) = e^{\int_0^t [D_0 + D_1(B_0(t) + z(1 - B_0(t)))]dx}$.

**Remark.** Let consider the service policy when we do not distinguish the customers of the same batch, i.e. all they have the same serve time. The arrival process is a MMAP with characteristic matrices $D_0$, $D(n) = \sum_{h_1 + h_2 + \ldots + h_K = n} D_\mathbf{h}$, where matrix $D(n)$ correspond to arrival of the batch with $n$ customers. In this case MMAP transforms to ordinary BMAP [36,37] with characteristic matrices $\{D_0, D(n), n > 0\}$.

The service time of customers $\gamma$ has general distribution $B(t)$ and mean value $\bar{\gamma}_1$. Thus for this model PGF we get

$$P(z,t) = e^{\int_0^t [D_0 + \sum_{n=1}^{\infty} D(n)(B(x) + z(1 - B(x)))^n]dx} = e^{\int_0^t [D_0 + D(B(x) + z(1 - B(x)))]dx},$$

where $D(z) = \sum_{n=1}^{\infty} D(n)(B(x) + z(1 - B(x)))^n$ is a rate GM of BMAP.

## 5. MMAP|G|∞ model with catastrophes

Let consider the general homogeneous Markovian model under influence of SMP generated catastrophes. As in [17], after every transition of environmental SMP the model jumps into the special state, let say 0-state, and then works from that state. When the SMP is in $i$ state all parameters of the model are related to that state: DF of inter-arrival time of customers, DF and rates of service time of customers, their resource vectors. Let $\bar{P}(\mathbf{n},t,i)$ and $P(\mathbf{n},t,i)$ defined the probabilities of having in the model $\mathbf{n} = (n_1, n_2, \ldots, n_k)$ customers at moment $t$ when environmental SMP is in state $i$, for the models without catastrophes and with catastrophes, resp. The following theorem gives the connection between these two models probabilities,

**Theorem 5.** The probabilities $P(\mathbf{n},t,i)$ satisfy the following integral equations

$$P(\mathbf{n},t,i) = (1-F_i(t))\bar{P}(\mathbf{n},t,i) + \sum_{j \in S} \int_0^t P(\mathbf{n},t-u,j)dQ_{ij}(u). \quad i \in S. \tag{12}$$

The solution of (12) equations can be found

$$P(\mathbf{n},t,i) = (1-F_i(t))\bar{P}(\mathbf{n},t,i) + \sum_{j \in S} \int_0^t (1-F_j(t-u))\bar{P}(\mathbf{n},t-u,j)dH_{ij}(u). \quad i \in S, \tag{13}$$

where $F(t) = \{F_i(t), i \in S\}$ is a sojourn time distribution vector of SMP: $F_i(t) = \sum_{j \in S} Q_{ij}(t), i \in S$.

$H(t) = \| H_{ij}(t) \|$ is a renewal matrix of SMP which components satisfy the following equations

$$H_{ij}(t) = 1 - F_i(t) + \sum_{k \in S} \int_0^t H_{kj}(t-u)dQ_{ik}(u), \quad i,j \in S. \tag{14}$$

The proof can be done by using standard renewal arguments (see for example [38]).

Let $\tilde{P}(z,t,i)$ and $\bar{\tilde{P}}(z,t,i)$ are the PGFs of $P(\mathbf{n},t,i)$ and $\bar{P}(\mathbf{n},t,i)$ respectively.

**Theorem 6.** The PGF $\tilde{P}(z,t,i)$ satisfy the following integral equations

$$\tilde{P}(z,t,i) = (1-F_i(t))\bar{\tilde{P}}(z,t,i) + \sum_{j \in S} \int_0^t \tilde{P}(z,t-u,j)dQ_{ij}(u), \quad i \in S, \tag{15}$$

which solution is

$$\tilde{P}(z,t,i) = (1-F_i(t))\bar{\tilde{P}}(z,t,i) + \sum_{j \in S} \int_0^t (1-F_j(t-u))\bar{\tilde{P}}(z,t-u,j)dH_{ij}(u), \quad i \in S, \tag{16}$$

The proof can be done by using standard renewal arguments or by CMM (e.g. see [38]).

**Theorem 7.** The limiting distributions of $P(\mathbf{n})$ and $\tilde{P}(z)$ are given by

$$P(\mathbf{n}) = \lim_{t \to \infty} P(\mathbf{n},t,i) = \sum_{j \in S} \frac{q_j}{\bar{\eta}_j} \int_0^\infty (1-F_j(u))\bar{P}(\mathbf{n},u,j)du, \quad i \in S, \tag{17}$$

$$\tilde{P}(z) = \lim_{t \to \infty} \tilde{P}(z,t,i) = \sum_{j \in S} \frac{q_j}{\bar{\eta}_j} \int_0^\infty (1-F_j(u))\bar{\tilde{P}}(z,u,j)du, \quad i \in S, \tag{18}$$

where $\bar{\eta}_i = \int_0^\infty (1-F_i(u))du$, $q_i = \frac{\bar{\eta}_i \rho_i}{\sum_{r \in S} \bar{\eta}_r \rho_r}$, $\sum_{r \in S} q_r = 1$, $\rho_i = \sum_{r \in S} p_{ri}\rho_r$, $\sum_{r \in S} \rho_r = 1$, $p_{ri} = Q_{ri}(\infty)$, $r,i \in S$.

Let $\hat{f}(s)$ denote the Laplace Transformation (LT) of a function $f(x)$, $\hat{f}(s) = \int_0^\infty e^{-su} f(u)du$.

The transient and stationary $k$-th order moments $L_{kr}(t)$ and $L_{kr}$ of $P(\mathbf{n},t,i)$ can be found from (15) – (18).

$$L_{1r}(t) = \frac{\partial \tilde{P}(z,t,i)}{\partial z_r}\bigg|_{z_1=z_2=...=z_k=1}, \quad L_{2r}(t) = \frac{\partial^2 \tilde{P}(z,t,i)}{\partial z_r^2}\bigg|_{z_1=z_2=...=z_k=1}, \quad L_{1r} = \lim_{t\to\infty} L_{1r}(t), \quad L_{2r} = \lim_{t\to\infty} L_{2r}(t), \quad r \in S. \tag{19}$$

**Corollary.** When $F_i(t) = 1 - e^{-\nu_i t}, i \in S$ then for LT of $P(\mathbf{n},t,i)$, PGF $\tilde{P}(z,t,i)$ and their limiting values we get

$$\hat{P}(\mathbf{n},s,i) = \hat{\bar{P}}(\mathbf{n},s+\nu_i,i) + \frac{1}{s}\sum_{j \in S} \hat{\bar{P}}(\mathbf{n},s+\nu_j,j)\hat{H}_{ij}(s), \quad i \in S, \tag{20}$$

$$\hat{\tilde{P}}(z,s,i) = \hat{\bar{\tilde{P}}}(z,s+\nu_i,i) + \frac{1}{s}\sum_{j \in S} \hat{\bar{\tilde{P}}}(z,s+\nu_j,j)\hat{H}_{ij}(s), \quad i \in S, \tag{21}$$

$$\tilde{P}(z) = \sum_{i \in S} q_i \nu_i \hat{\bar{\tilde{P}}}(z,\nu_i,i), \quad P(\mathbf{n}) = \sum_{i \in S} q_i \nu_i \hat{\bar{P}}(\mathbf{n},\nu_i,i). \tag{22}$$

Let consider the infinite-server models MMAP|G|∞ with catastrophes. In this model the environmental SMP can be considered as a catastrophes process. After every transition of this SMP all customers in the model are flushed out instantly, the model jumps into empty state, and then continues its work from this state. If SMP is in state $i$ at moment $t$ all parameters of

the model - arrival process and service time distributions of customers, are function of $i$. Let $L(\mathbf{n},\mathbf{x},t,i)$ is a matrix, $L(\mathbf{n},\mathbf{x},t,i) = \|L_{jr}(\mathbf{n},\mathbf{x},t,i)\|$, where $L_{jr}(\mathbf{n},\mathbf{x},t,i)$ is the probability of event "in the model there are $\mathbf{n} = (n_1, n_2, ..., n_K)$ customers at moment $t$, PP $J(t)$ is in the phase $j$, total accumulated resource is $\mathbf{x}$ and SMP $\xi(t)$ is in state $i$ under condition that at initial moment $t = 0$ the model was empty, and PP $J(0)$ was in phase $r \in S$ ",

$$L_{jr}(\mathbf{n},\mathbf{x},t,i) = P(\mathbf{N}_i^s(t) = \mathbf{n}, \boldsymbol{\alpha}(t) \le \mathbf{x}, J(t) = j, \xi(t) = i \mid \mathbf{N}_i^s(0) = \mathbf{0}, J(0) = r).$$

Let $L(\mathbf{n}, s, t, i)$ be the LST of $L(\mathbf{n},\mathbf{x},t,i)$.

**Theorem 8.** The probabilities $L(\mathbf{n}, s, t, i)$ and $L(\mathbf{n}, s, t)$ satisfy the following renewal equations

$$L(\mathbf{n}, s, t, i) = (1 - F_i(t)) \bar{P}(\mathbf{n}, s, t, i) + \sum_{j \in S} \int_0^t L(\mathbf{n}, s, t - u, j) dQ_{ij}(u), \qquad (23)$$

Which solutions are given by

$$L(\mathbf{n}, s, t, i) = (1 - F_i(t)) \bar{P}(\mathbf{n}, s, t, i) + \sum_{j \in S} \int_0^t (1 - F_j(t - u)) \bar{P}(\mathbf{n}, s, t - u, j) dH_{ij}(u), \qquad (24)$$

$$L(\mathbf{n}, s, t) = \sum_{i \in S} p_i^0 (1 - F_i(t)) \bar{P}(\mathbf{n}, s, t, i) + \sum_{j \in S} \int_0^t (1 - F_j(t - u)) \bar{P}(\mathbf{n}, s, t - u, j) dH_j(u),$$

where $H_j(t)$ is a renewal function of SMP $\xi(t)$: $H_j(t) = \sum_{k \in S} p_k^0 H_{kj}(t)$.

The theorem can be proved by using properties of renewal arguments [38].

If SMP $\xi(t), t \ge 0$ is an irreducible, ergodic process then for $L(\mathbf{n}, s, t)$ when $t$ is tending toward $+\infty$ by means *key renewal theorem* from (18) we derive following asymptotic result

$$L(\mathbf{n}, s) = \lim_{t \to \infty} L(\mathbf{n}, s, t) = \sum_{i \in S} \frac{q_i}{\eta_i} \int_0^\infty (1 - F_i(u)) \bar{P}(\mathbf{n}, s, u, i) du. \qquad (25)$$

Let $\tilde{L}(z, s, t)$ be PGF of $L(\mathbf{n}, s, t)$, $\tilde{L}(z, s, t) = \sum_{n=0}^\infty z^n L(\mathbf{n}, s, t)$, and $\tilde{L}(z, s) = \lim_{t \to \infty} \tilde{L}(z, s, t)$.

Thus for $\tilde{L}(z, s, t)$ and $\tilde{L}(z, s)$ we derive

$$\tilde{L}(z, s, t) = \sum_{i \in S} p_i^0 (1 - F_i(t)) \tilde{\bar{P}}(z, s, t, i) + \sum_{j \in S} \int_0^t (1 - F_j(t - u)) \tilde{\bar{P}}(z, s, t - u, j) dH_j(u), \qquad (26)$$

$$\tilde{L}(z, s) = \sum_{i \in S} \frac{q_i}{\eta_i} \int_0^\infty (1 - F_i(u)) \tilde{\bar{P}}(z, s, u, i) du.$$

By substitution the expression for $\tilde{\bar{P}}(z, s, t, i)$ into (25) for PGF of joint distribution of number of busy servers and total accumulated resources in the model at moment $t$ given that at $t = 0$ SMP was in state $i$ for the model with catastrophes and its limiting value we get

**Theorem 9**. The PGFs $\tilde{L}(z, s, t, i)$, $\tilde{L}(z, s, t)$ and its limiting value $\tilde{L}(z, s)$ are given by

$$\tilde{L}(z, s, t, i) = e^{\int_0^t [D_0(i) + \tilde{S}_i(z, s, u)] du} (1 - F_i(t)) + \sum_{r \in S} \int_0^t e^{\int_0^{t-u} [D_0(r) + \tilde{S}_r(z, s, x)] dx} dQ_{ir}(u), \; i \in S, \qquad (27)$$

which solutions are

$$\tilde{L}(z, s, t, i) = e^{\int_0^t [D_0(i) + \tilde{S}_i(z, s, u)] du} (1 - F_i(t)) + \sum_{r \in S} \int_0^t (1 - F_r(t - u)) e^{\int_0^{t-u} [D_0(r) + \tilde{S}_r(z, s, x)] dx} dH_{ir}(u), \quad i \in S, \qquad (28)$$

$$\tilde{L}(z, s, t) = \sum_{i \in S} p_i^0 e^{\int_0^t [D_0(i) + \tilde{S}_i(z, s, u)] du} (1 - F_i(t)) + \sum_{r \in S} \int_0^t (1 - F_r(t - u)) e^{\int_0^{t-u} [D_0(r) + \tilde{S}_r(z, s, x)] dx} dH_r(u),$$

$$\tilde{L}(z, s) = \lim_{t \to \infty} \tilde{L}(z, s, t, r) = \sum_{i \in S} \frac{q_i}{\eta_i} \int_0^\infty (1 - F_i(x)) e^{\int_0^x [D_0(i) + \tilde{S}_i(z, s, u)] du} dx, \; r \in S. \qquad (29)$$

**Corollary.** When $F_i(t) = 1 - e^{-v_i}$, $i \in S$ then for LT of PGF $\tilde{L}(z,s,t,i)$ and their limiting values we get

$$\tilde{L}(z,s,s,i) = \tilde{\bar{P}}(z,s,s+v_i,i) + \frac{1}{s}\sum_{r \in S} \tilde{\bar{P}}(z,s,s+v_r,r)\hat{H}_{ir}(s). \quad i \in S \tag{30}$$

$$\tilde{L}(z,s) = \sum_{r \in S} \tilde{\bar{P}}(z,s,v_r,r)q_r v_r. \tag{31}$$

If SMP is a simple renewal process with exponential distributed renewal time $F(t) = 1 - e^{-vt}$ then by (31) we get the results for homogeneous model [20]. For example, for $\tilde{L}(z,s)$ and $\tilde{L}(z,s,s)$ we obtain

$$\hat{\tilde{L}}(z,s,s) = \left(1 + \frac{v}{s}\right)\tilde{\bar{P}}(z,s,s+v), \quad \tilde{L}(z,s) = v\tilde{\bar{P}}(z,s,v).$$

## 6. Performance measures of the model.

Let $\omega_{1r}(t)$, $\omega_{1r}$ denote the transient and steady state mean of queue length of type $r$ customers. Then for $\omega_{1r}(t)$ and $\omega_{1r}$ we get

$$\omega_{1r}(t) = \lim_{s_1 \to 0} \omega_{1r}(s_1,t), \quad \omega_{1r}(s_1,t) = \left.\frac{\partial \tilde{L}(z_1,s_1,t)}{\partial z_{1r}}\right|_{z_{1r}=1, z_{11}=z_{12}=...=z_{1k}=1},$$

$$\omega_{1r} = \lim_{s_1 \to 0} \omega_{1r}(s_1), \quad \omega_{1r}(s_1) = \left.\frac{\partial \tilde{L}(z_1,s_1)}{\partial z_{1r}}\right|_{z_{1r}=1, z_{11}=z_{12}=...=z_{1k}=1},$$

$$\omega_{1r}(s_1,t) = \sum_{i \in S} p_i^0(1 - F_i(t))\bar{\omega}_{1r}(s_1,t,i) + \sum_{j \in S} \int_0^t (1 - F_j(t-u))\bar{\omega}_{1r}(s_1, t-u, j)dH_j(u),$$

$$\omega_{1r}(s_1) = \sum_{j \in S} \frac{q_j}{\bar{\eta}_j} \int_0^\infty (1 - F_j(u))\bar{\omega}_{1r}(s_1, u, j)du. \tag{32}$$

Where $\bar{\omega}_{1r}(s_1,t,i)$ is a transient mean queue length of $r$ type customers of the model without catastrophes when the SMP is in state $i$.

Let $\delta_r(t)$, $\delta_r$, $r = 1,2,...,K$, be the transient and steady-state mean values of accumulated type $r$ resources in the model and $\delta$ be a total accumulated resources in the model.

$$\delta_r(t) = \pi \bar{\delta}_r(t)e, \quad \delta_r(t) = \lim_{s_1 \to 0} \left.\frac{\partial \tilde{L}(z_1,s_1,t)}{\partial s_{1r}}\right|_{z_{11}=z_{12}=...=z_{1k}=1},$$

$$\delta_r(t) = \sum_{i \in S} p_i^0(1 - F_i(t))\bar{\delta}_r(t,i) + \sum_{j \in S} \int_0^t (1 - F_j(t-u))\bar{\delta}_r(t-u, j)dH_j(u),$$

$$\delta_r = \lim_{t \to \infty} \delta_r(t), \quad \delta_r = \sum_{j \in S} \frac{q_j}{\bar{\eta}_j} \lambda_{jr}\bar{c}_{1r}(j) \int_0^\infty \int_0^u (1 - F_j(u))(1 - B_{jr}(x))dxdu, \tag{33}$$

where $\bar{\delta}_r(t,j) = \lambda_{jr}\bar{c}_{1r}(j)\int_0^t (1 - B_{jr}(x))dx$, $\bar{c}_{1r}(j)$ is a mean value of DF $C_{jr}(x)$.

$$\delta = \sum_{r=1}^K \delta_r = \sum_{r=1}^K \sum_{j \in S} \frac{q_j}{\bar{\eta}_j}\lambda_{jr}\bar{c}_{1r}(j)\int_0^\infty \int_0^u (1 - F_j(u))(1 - B_{jr}(x))dxdu.$$

If $L_{losr}$ denote the steady state mean number of destroyed type $r$ customers, then

$$L_{losr} = \lim_{s_1 \to 0} \pi L_{losr}(s_1)e, \tag{34}$$

where $L_{losr}(s_1) = \sum_{j \in S} \frac{q_j}{\bar{\eta}_j}\int_0^\infty \bar{\omega}_{1r}(s_1, u, j)dF_j(u)$.

$$L_{losr} = \sum_{j \in S} \frac{q_j}{\bar{\eta}_j} \lambda_{jr} \int_0^\infty \int_0^u (1 - B_{jr}(x)) dx dF_j(u). \tag{35}$$

If $L_{los}$ is the steady-state total mean number of destroyed customers of all types, then

$$L_{los} = \sum_{r=1}^K L_{losr} = \sum_{r=1}^K \sum_{j \in S} \frac{q_j}{\bar{\eta}_j} \lambda_{jr} \int_0^\infty \int_0^u (1 - B_{jr}(x)) dx dF_j(u). \tag{36}$$

Let $L_{qr}$ and $L_q$ be the steady state mean number of type $r$ and all types customers in the model. Then

$$L_{qr} = \pi \tilde{\omega}_{1r} e = \sum_{j \in S} \frac{q_j}{\bar{\eta}_j} \int_0^\infty (1 - F_j(u)) \pi \tilde{\bar{\omega}}_{1r}(u, j) e du = \sum_{j \in S} \frac{q_j}{\bar{\eta}_j} \lambda_{jr} \int_0^\infty \int_0^u (1 - F_j(u))(1 - B_{jr}(x)) dx du,$$

$$L_q = \sum_{r=1}^K L_{qr} = \sum_{r=1}^K \sum_{j \in S} \frac{q_j}{\bar{\eta}_j} \lambda_{jr} \int_0^\infty \int_0^u (1 - F_j(u))(1 - B_{jr}(x)) dx du. \tag{37}$$

Suppose that MMAP is defined by following matrices $D_0(i) = -\alpha_i I$, $D_h(i) = \alpha_i p(h_1, h_2, ..., h_K) I;\ i \in S,\ h \in C^0$, where $I$ is an identity matrix. Then for $\bar{P}(n, t, i)$, $L(n, t)$, $L(n)$, and $L_{losr}$ we obtain

$$\tilde{D}_i(z, u, t) = \sum_{n=0}^\infty p(n_1, n_2, ..., n_K) \prod_{r=1}^K [z_{ri}(1 - G_{ri}(t - u)) + G_{ri}(t - u)]^{n_r},$$

$$\tilde{\bar{P}}(z, t, i) = e^{-\alpha_i \int_0^t \{1 - \tilde{D}_i(z, u, t)\} du}, \tag{38}$$

$$\tilde{L}(z, t) = \sum_{j \in S} p_i^0 (1 - F_i(t)) e^{-\alpha_i \int_0^t \{1 - \tilde{D}_i(z, u, t)\} du} + \sum_{j \in S} \int_0^t (1 - F_j(t - u)) e^{-\alpha_j \int_0^{t-u} \{1 - \tilde{D}_j(z, u, t - u)\} du} dH_j(u),$$

$$\tilde{L}(z) = \sum_{i \in S} \frac{q_i}{\bar{\eta}_i} \int_0^\infty (1 - F_i(u)) e^{-\alpha_i \int_0^t \{1 - \tilde{D}_i(z, u, t)\} du} du.$$

$$L_{los} = \sum_{r=1}^K \sum_{j \in S} \frac{q_j}{\bar{\eta}_j} \lambda_{jr} \int_0^\infty \int_0^u (1 - B_{jr}(x)) dx dF_j(u),$$

where $\lambda_{ir} = \alpha_i \sum_{n_1=0}^\infty ... \sum_{n_K=0}^\infty n_r p(n_1, n_2, ..., n_K)$.

The (38) is a known result for Mr|Gr|∞ model (se for example [13] formula (5)).

### Conclusion

We consider infinite-server MMAP$_k$|G$_k$|∞ queue in SM random environment, marked MAP arrival of customers, random resource vector of each type of customers and catastrophes. The joint PGF of transient and stationary distributions of number of busy servers and numbers of served customers, as well as the LT of joint distributions of total accumulated resource in the model and total accumulated resource vectors of served customers during time interval $[0, t)$ are found. The renewal equations for transient and stationary PGF of queue sizes and resource vectors of different type of customers are found for MMAP$_k$|G$_k$|∞ and BG$_k$|G$_k$|∞ queue models. For homogeneous Markov model in SM random environment and catastrophes the transient and limiting distributions renewal equations and their solutions are found. All results are obtained using CMM and renewal process methods.

The obtained results may be applied for computer system and network performance metrics evaluation, as well as for design of optimal strategies of resource management of a wide class of subsystems of New Generation Networks, whereas the queue MMAP$_r$|G$_r$|∞ may be used as a model of these subsystems.